# Enhancing keV high harmonic signals generated by long-wave infrared lasers

ZENGHU CHANG

*Institute for the Frontier of Attosecond Science and Technology, CREOL and Department of Physics, University of Central Florida, 4111 Libra Drive, PS430, Orlando, FL 32816, USA*
*Zenghu.Chang@ucf.edu*

**Abstract:** It is demonstrated by single-atom simulations that X-ray signals in the 3.4 to 4 keV region from an 8 micron laser driven high harmonic generation can be increased by more than two orders of magnitude when a single-cycle pulse centered at 800 nm is added. The ionization probability of a helium atom by the two-pulse field is set to $4.56 \times 10^{-5}$, which is needed for balancing the index of refraction of free electrons with that of neutral helium atoms to achieve phase matching.

## 1. Introduction

Since first demonstrated in 2001, table-top attosecond light sources have relied on high harmonic generation (HHG) in noble gases interacting with near-infrared (NIR) Ti:Sapphire lasers centered at 800 nm [1, 2]. The attosecond spectrum with sufficient flux for transient absorption and streaking experiments is limited to the extreme ultraviolet region (10 to 150 eV). It was demonstrated that the cutoff photon energy of high harmonics can be dramatically extended by driving the non-perturbative nonlinear process with long wavelength lasers [3]. In the last decade, mJ-level driving lasers centered in the 1.3 to 4 μm range have been developed for generating X-ray pulses in the water-window region (282 to 533 eV) and beyond [4-7]. Long-wave infrared (LWIR) lasers centered at 8 μm are being designed [8] to develop attosecond X-ray sources covering the photon energy range of 1 to 5 keV.

It is observed experimentally that the conversion efficiency of the high harmonic generation may decrease when the driving laser wavelength is increased, which was attributed to effects of quantum diffusion of electron wave packets[3]. This topic has been studied both theoretically and experimentally[9, 10]. As the driving laser wavelength moves from near infrared to mid-wave (3–8 μm) and long-wave (8–15 μm), innovative schemes are needed to achieve useful photon flux. Multi-color laser fields have also been extensively investigated to increase the high harmonic yield and to extend the cutoff [11, 12]. The high harmonic generation signal in the plateau region can be enhanced by driving the process with a combination of a strong infrared (IR) and a weak attosecond extreme ultraviolet (XUV) pulse train [13-15] or single isolated sub-femtosecond vacuum ultraviolet (VUV) pulses [16, 17]. Since the duration of the XUV/VUV pulse is much shorter than the IR optical period, the particular ionization time can be fixed by properly setting the time delay between the two pulses. The typical intensities of the IR and XUV (or VUV) are $10^{14}$ W/cm$^2$ and $10^{13}$ W/cm$^2$ respectively. Numerical simulation also shows that a few-cycle 400 nm pulse added to a stronger 2000 nm pulse can enhance the high harmonic signal in the 100 to 400 eV region [18]. In addition, it has been demonstrated theoretically that the cutoff of the high harmonic generated by an intense 800 nm pulse can be extended by a 10 times weaker 8 micron pulse to generate isolated XUV attosecond pulses in the 23 to 93 eV range[19]. Here we consider the case where high harmonic generation for producing keV X-rays is driven by a long-wave infrared laser. It is found that the keV X-ray signal can be significantly enhanced by adding a single-cycle carrier-envelope phase (CEP) stable NIR laser centered at 800 nm with peak intensities that are a few times less that of the LWIR.

## 2. Ionization probability for phase matching of high harmonic generation

High harmonic generation in helium atoms is studied in this work. The index of refraction of helium gas has been measured at 633 nm[20] and 10.7 micron [21]. However, its value at 8 micron is hard to find in literatures and is estimated by assuming that the frequency dependence of the index of refraction of neutral helium can be approximately expressed by a single-resonance classical electron oscillator equation [22]

$$n_a(\omega_L) = 1 + \frac{e^2}{2\varepsilon_0 m_e} \frac{N_a}{\omega_r^2 - \omega_L^2}, \qquad (1)$$

where $e$ and $m_e$ are the charge and mass of an electron respectively. $\varepsilon_0$ is the permittivity of free space. $N_a$ is the atomic number density of the unionized atom. $\omega_L$ and $\omega_r$ are the angular frequency of the laser and the resonant frequency of the atom respectively. The resonant photon energy $\hbar\omega_r = 22.963$ eV can be obtained by using the measured index of refraction value at 10.7 micron.

The ionization probability, $p_{pm}$, of the helium atom is fixed when the high harmonic signals generated by a pulse centered at 8 micron alone are compared with that produced by a combination of 8 micron and 800 nm pulses. To achieve phase matching in high harmonic generation[23], the value of $p_{pm}$ is chosen so that the index of refraction of the unionized portion of the gas at 8 micron cancels that of the ionized portion, which can be calculated by[22]

$$p_{pm}(\omega_L) = \left(\frac{\hbar\omega_L}{\hbar\omega_r}\right)^2. \qquad (2)$$

It results in $p_m = 4.556 \times 10^{-5}$.

The high harmonic signals generated by a driving laser field centered at 8.0 µm alone serves as the reference to demonstrate the effects of the additional NIR pulse. The LWIR field and the ionization probability of a helium atom is shown in Fig. 1(a).

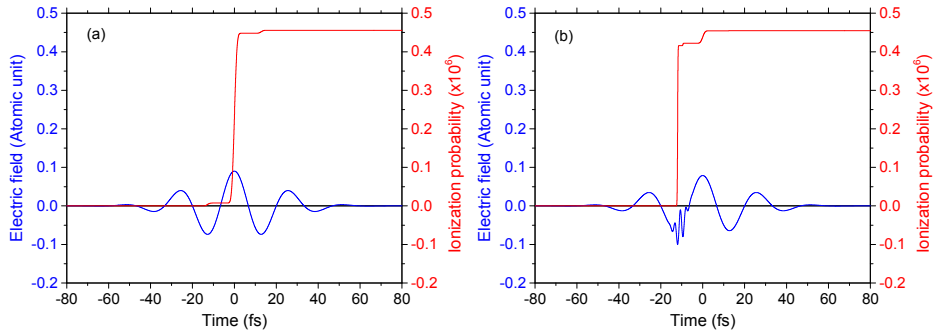

Fig. 1. (a) A 34 fs LWIR driving laser field centered at 8.0 µm (blue line) and the ionization probability of a helium atom (red line). The peak intensity is $2.86 \times 10^{14}$ W/cm$^2$. The carrier envelope phase is 0 rad. (b) A 3.4 fs NIR pulse centered at 800 nm is added to a LWIR pulse whose intensity is $2.18 \times 10^{14}$ W/cm$^2$. The intensity of the NIR pulse is 1/3 that of the LWIR pulse. The CEP of the NIR pulse is -0.982 rad. The peak of the NIR pulse envelope is 1.067 fs ahead of the LWIR pulse.

The FWHM duration, and carrier envelope phase (CEP) of the LWIR pulse are 34 fs, and 0 rad respectively, such a short pulse allows generation of isolated attosecond pulses with the amplitude gating[24]. When the HHG is driven by the 8 micron pulse only, the peak intensity of the laser is set at $2.86\times10^{14}$ W/cm$^2$ so that ionization probability at the end of the pulse is $4.56\times10^{-5}$.

When a 3.4 fs NIR pulse centered 800 nm is added to the 8 micron pulse to enhance the HHG signal, the peak intensity of the latter needs to be reduced to maintain the same final ionization probability. The combined electric field is illustrated in Fig. 1 (b). The intensities of the LWIR and NIR pulses are $2.18\times10^{14}$ W/cm$^2$ and $7.27\times10^{13}$ W/cm$^2$ respectively. The carrier-envelope phase of the NIR pulse is -0.982 rad. The peak of the NIR pulse envelope is 1.067 fs ahead of the LWIR pulse. The CEP and delay values are chosen to optimize the HHG enhancement as described in section 3. Since the 800 nm pulse duration, 3.4 fs, is shorter than a quarter of optical cycle of the 8 micron light, 5 fs, the tunneling ionization induced by the NIR field is localized that may enhance the emission of a single X-ray burst.

## 3. Numerical simulations of X-ray signal enhancement

Numerical simulations based on the Strong Field Approximation of high harmonic generation [25] have been performed to demonstrate the enhancement of keV X-ray signals by adding a NIR laser pulse to a LWIR field. The time domain dipole moment of a single helium interacting with a LWIR laser pulse alone or a combined LWIR and NIR field was calculated using the open-source code, HHGmax [26]. The dipole matrix element is hydrogen-like and the ionization potential of the atom is 24.59 eV.

In Figure 2(a), the power spectrum of the high harmonics generated with the LWIR pulse alone (yellow line) is plotted together with that from the two-pulse driving field (blue line). The laser parameters are identical to that for Fig. 1. The X-ray intensity in the 3.4 to 4 keV region is increased by more than two orders of magnitude when the 800 nm pulse is added. The cutoff photon energy reaches 5.4 keV for the 8 micron laser generated harmonics, which agrees with the semi-classical cutoff law. The cutoff is reduced to 4.1 keV when the high harmonic generation is driven by the two-pulse field because LWIR intensity is reduced. The Fourier transforms of the spectra of the last plateau before the cutoff are shown in Fig. 2(b). In addition to the electric amplitude increase, the X-ray pulse duration is shortened by about a factor of two when the 800 nm pulse is added, which is due the suppression of the long quantum trajectory.

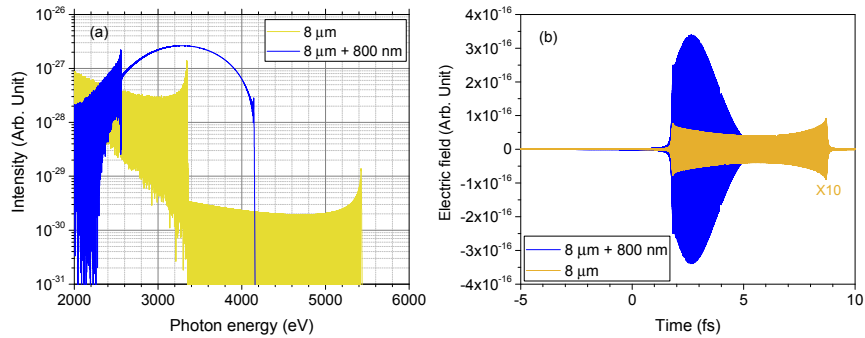

Fig. 2. (a) Power spectra of the high harmonics generated with a LWIR pulse alone (yellow line) and with a LWIR pulse combined with a NIR pulse. The laser parameters are explained in Fig. 1. (b) X-ray pulses corresponding to the spectra in the plateau adjacent to the cutoff.

Both the X-ray spectral shape and intensity depend strongly on the carrier-envelope phase of the 800 nm pulses, as shown in Fig. 3 (a). They are also sensitive to the CEP of the

LWIR pulse, as well as the time delay between the two pulses. Therefore, the three parameters must be stabilized in experiments. CEP stable few-cycle lasers centered at 800 nm have been the working horse for generating single isolated attosecond pulses[2]. When the seed pulses of the 8 micron laser are generated with the 800 nm laser such as the architecture described in [8], the precision synchronization of the two pulses can be accomplished optically and the CEP of the 8 micron pulse can be stabilized.

At a given ionization probability, the intensity of the 800 nm pulse can be chosen by finding an acceptable compromise between the cutoff photon energy and the X-ray signal strength. The comparison of high harmonic spectra generated with the combined fields at two different NIR intensities is shown in Fig. 3(b). When the intensity of the 800 nm pulse is one fifth of the 8 micron pulse, the cutoff reaches a higher value than that when the intensity ratio is one to three, the enhancement of the X-ray intensity is less (but still more than two orders of magnitude around 4 keV). The requirement of the intensity of the quasi-single-cycle 800 nm pulse can be met with Ti:Sapphire lasers followed by a hollow-core fiber compressor, which seeds the 8 micron laser[8].

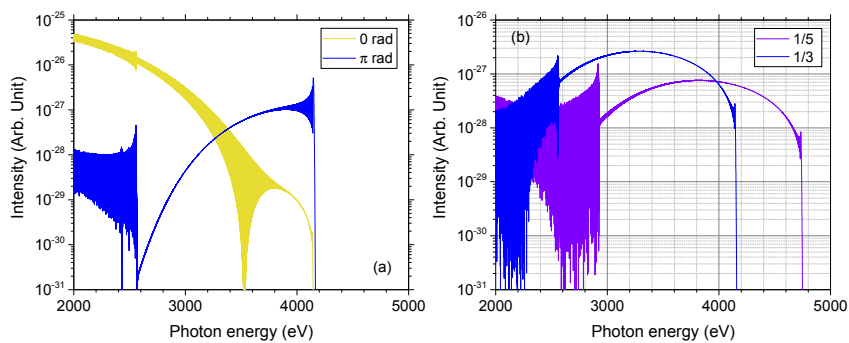

Fig. 3. High harmonic spectra generated with the combined 8 micron and 800 nm pulses. (a) The CEP of the 800 nm pulse is changed. (b) The intensity ratio between the 800 nm and 8 micron pulses is varied.

### 4. Summary

More than two orders of magnitude enhancement of high harmonic signals in the keV photon energy region is achievable by combining LWIR and NIR pulses with comparable intensities as demonstrated by numerical simulations. The long optical period of the 8 micron laser (26.6 fs) allows the control of ionization within a quarter of a cycle by a 3.4 fs pulse centered at 800 nm. With the recent progress on high energy picosecond 2 µm pump lasers, it is expected that Optical Parametric Chirped Pulse Amplifiers centered at 8 µm may soon become available. The single atom simulations in this study are done at the phase matching ionization probability, further investigation will be conducted to study the effects of propagation on the X-ray yield.

**Funding.** Air Force Office of Scientific Research (AFOSR) (FA9550-15-1-0037, FA9550-16-1-0013); Army Research Office (ARO) (W911NF-14-1-0383, W911NF-19-1-0224); Defense Advanced Research Projects Agency (DARPA) Topological Excitations in Electronics (TEE) (D18AC00011); National Science Foundation (NSF) (1806584).